\PassOptionsToPackage{table,xcdraw}{xcolor}
\documentclass[sigconf, submit]{acmart}

\renewcommand\footnotetextcopyrightpermission[1]{} 
\usepackage{amsmath}
\usepackage[table,xcdraw]{xcolor}

\usepackage{xcolor}
\usepackage{adjustbox}
\definecolor{citegreen}{HTML}{458B00}
\usepackage{multirow}
\usepackage{makecell}
\usepackage[utf8]{inputenc}

\usepackage{longtable}
\usepackage{pgfplots} 
\usepackage{tikz,pgfplots,pgfplotstable}
\pgfplotsset{compat=1.8}
\PassOptionsToPackage{hyphens}{url}\usepackage{hyperref}
\usepackage{graphicx}
\hypersetup{
   colorlinks=true,
   citecolor=citegreen
}
\usepackage{caption}
\usepackage{subcaption}
\usepackage{booktabs}

\usepackage{etoolbox}

\AtBeginDocument{%
  \providecommand\BibTeX{{%
    \normalfont B\kern-0.5em{\scshape i\kern-0.25em b}\kern-0.8em\TeX}}}

\begin{document}
\title{An Empirical Analysis of SMS Scam Detection Systems}

\author{Muhammad Salman}
\affiliation{%
  \institution{Macquarie University}
  \city{Sydney}
  \country{Australia}}
\email{muhammad.salman2@students.mq.edu.au}

\author{Muhammad Ikram}
\affiliation{%
  \institution{Macquarie University}
  \city{Sydney}
  \country{Australia}}
\email{muhammad.ikram@mq.edu.au}

\author{Mohamed Ali Kaafar}
\affiliation{%
  \institution{Macquarie University}
  \city{Sydney}
  \country{Australia}}
\email{dali.kaafar@mq.edu.au}


\sloppy

\begin{abstract} 
The short message service (SMS) was introduced a generation ago to the mobile phone users. They make up the world's oldest large-scale network, with billions of users and therefore attracts a lot of fraud. Due to the convergence of mobile network with internet, SMS based scams can potentially compromise the security of internet services as well. In this study, we present a new SMS scam dataset consisting of 153,551 SMSes. This dataset that we will release publicly for research purposes represents the largest publicly-available SMS scam dataset. We evaluate and compare the performance achieved by several established machine learning methods on the new dataset, ranging from shallow machine learning approaches to deep neural networks to syntactic and semantic feature models. We then study the existing models from an adversarial viewpoint by assessing its robustness against different level of adversarial manipulation. This perspective consolidates the current state of the art in SMS Spam filtering, highlights the limitations and the opportunities to improve the existing approaches.
\end{abstract}

\maketitle


\section{Introduction}
\label{sec:intro}


SMS scams have reached alarming volumes in recent years. In the US in 2021, an estimated USD 86 million was lost to SMS scammers (reported by Federal Trade Commission\footnote{https://www.fcc.gov/covid-19-text-scams}). Similarly, the Australian Competition and Consumer Commission (ACCC)'s ScamWatch body\footnote{https://www.scamwatch.gov.au/scam-statistics} reported a near-doubling of annual losses between 2020 (AUD 175 millions) and 2021 (AUD 323 millions). In 2021, there were 67,180 SMS fraud reports, up from 32,337 reported in 2020: more than 8,835 SMS scams were reported only in February 2022, the highest among all scam delivery methods. Despite almost two decades of research on SMS Spam detection \cite{buchanan2001investigating,zhang2004evaluation, wittel2004attacking, webb2005experimental, almeida2011contributions, almeida2013towards, gupta2018comparative,rojas2021using}, it continues to be a challenging and serious issue to our modern digital societies.

Researchers have developed various techniques to combat SMS Spam, and there are several surveys of these attempts \cite{rojas2021using,oswald2022spotspam,yerima2022semi}. However, unlike email spam detection, it is difficult to identify the appropriate system and directly compare the proposed systems in literature, since most of them used highly outdated and different datasets for evaluation. Although, traditional ML and deep learning (DL) methods have substantially improved Spam detection in SMS; however, scammers' dynamic nature frequently defeats security barriers. Recently, studies have revealed different adversarial attacks in the text domain, which could effectively evade various ML-based text analyzers \cite{wang2019towards,huq2020adversarial,rojas2021using,gao2018black}. We discovered that there are certain methodological issues with current SMS Spam detection research and it requires a lot more research. For example, we observe:
\begin{itemize}
\item The use of highly outdated and imbalanced datasets with only few hundreds of Spam messages
\item Lack of benchmark, which resulted in partial and segmented evaluation of literature
\item Lack of comprehensive quantification and comparison of different classes of ML such as DL, positive and unlabeled (PU) learning, traditional two class and one class ML etc. The same is observed for different set of syntactic and semantic features
\item Importantly, the literature overlook new challenges that arise as convergence of Internet and Telcom especially the new adversarial attempts of scammers to evade detection
\end{itemize}

The focus of this paper is to investigate the landscape of fraudulent SMS messages, existing methods for combating SMS scams, state-of-the-art research in SMS Spam defense, points out the required set of challenges and to analyse how current tactics may be abused by scammers and subsequently identify the requirements that are crucial to an acceptable solution. Therefore, we assess the performance and robustness of the SMS Spam models proposed in the literature on the new super dataset (shortly discussed in section \ref{sec:data}), based on the requirements of the security domain, which we elaborate upon in section \ref{sec:aml}, to determine the appropriateness of the proposed solutions. To our knowledge, a comprehensive experimental evaluation of the appropriateness of the previous research on SMS Spam from the security perspective has not been done before. To fill this important gap, this study will assist drive the development of effective/ successful SMS Spam defenses, as well as provide a framework for evaluating SMS Spam methods. The major contributions of our paper are:

{\bf Large scale Spam Dataset.} We make available a new real, public and large SMS Spam corpus from various sources over the previous decade, that is the largest labeled dataset as far as we are aware. We aggregate and characterise publicly available Spam dataset. Given that the benchmark dataset \cite{UCIMLRep} used for SMS Spam research is outdated and incomplete consisting only 5,574 messages with 13\% Spam and 87\% Ham (or legitimate) messages, it is not suitable for detecting the most recent, stealthy SMS Spam even using advanced machine or DL techniques. We propose data collection methods to crawl 55,686 SMS messages from Spam observatories such as ScamWatch\footnote{https://www.scamwatch.gov.au/} and Action Fraud\footnote{ https://www.actionfraud.police.uk/}. We labelled the dataset into Spam and non-Spam messages constituting 15,209 (27.31 \%) and 40,477 (72.69 \%) SMSes, respectively. 

{\bf Valuation of machine learning models.}
Usually, supervised text classification methods (like binary classification) or DL have been used to identify SMS Spam. However, given the imbalance nature of the SMS Spam datasets, we propose one-class learning (unsupervised learning) and positive unlabeled (PU) learning (semi-supervised learning) algorithms alongside other ML models. To the best of our knowledge, the notion of one-class learning and PU learning for SMS Spam detection was never explored earlier. We evaluate and compare the performance of PU learning and one-class learning models with both
traditional two class ML as well as with the state of the art neural network and DL
such as the latest transformers based architecture \cite{vaswani2017attention}. Furthermore, we analysed and compared the performance of several well-known ML methods in order to establish a good baseline for future comparisons. Additionally, we evaluated the ML models over the new dataset with different set of features/ word embeddings starting from Non-semantic Count based Vector space model (Count Vectorization, TF-IDF) to semantic Non Context-Based Vector Space Model (Word2Vec, fastText, GloVe) with static and dynamic modes and with semantic Context Based Vector Space Models (BERT, ELMo etc).

{\bf Robustness analysis of Spam models in the face of new challenges.}
To evade detection, Spammers may leverage obfuscation or perturbation methods \cite{morris2020textattack} to Smish \cite{salahdine2019social} or Punycode \cite{costello2003punycode} Spam SMS texts and embedded URLs. We evaluate the robustness of ML models against 4 levels of adversarial perturbations: charlevel, word-level, sentence-level, and multi-level attacks. To the best of our knowledge, such extensive adversarial assessment of the ML models on Spam SMS have not been performed before.

\section{Research Challenges and Related Work}

In the following, we discuss challenges and overview sate-of-the-art SMS Spam filtering and detection methods in literature.

\subsection{Challenges}
{\bf Availability of data.} The lack of updated, genuine, and publicly large SMS Spam dataset is a major concern. The SMS Spam filtering has a lack of good dataset sources and a dearth of diversity in the data used for evaluation of detection systems. Recency of the data is another challenge. The datasets used by researchers in the literature are extremely outdated and raises questions on its suitability in the present SMS scam landscapes due to non-representative of latest attacks (see table \ref{tab:consol}). Other concern is the message ambiguity due to short length), limited header information, presence of emojis and abbreviations in their text, therefore, established email Spam filters may have their performance seriously degraded when directly employed to dealing with mobile Spam. 

{\bf Lack of benchmark.} Different methods have been proposed by researchers to study SMS Spam detection and adversarial attacks in texts, but there is no benchmark \cite{shafi2017review,li2018textbugger}. Moreover, different datasets have been used by researchers in their work, making it difficult to compare these methods. Meanwhile, it also affects the selection of evaluation metrics. There is no exact statement about which metric measure is better in a situation and why it is more useful than others.

{\bf Robustness against new challenges and attacks.} Another challenge in the Spam detection is the dynamic behavior of scammers. They always try to find a way to deceive the Spam filters using different adversarial attacks. Moreover, due to the convergence of mobile network with internet, Spammers may likely adapt the techniques used for tricking users on internet.

{\bf Lack of strategy and collaboration} One of the less highlighted issue in effective tackling of Spam SMS is the lack of collaborations between researchers, end users, network providers, and industry as well as absence of a robust strategy for dealing with threats to the security of mobile users and Spam SMS detection.

\subsection{Related Work}
To effectively handle the threat posed by SMS Spam, several techniques have been proposed in the academic literature and white papers in the industry. However, none of these studies provides a complete picture of the SMS Spam problem, though there are resources that handle part of the problem or try to reduce the problem to a single dimension. Importantly, there is no previous systematic empirical analysis of SMS scam detection systems.

Abdulhamid et al.\cite{shafi2017review} presented a review of the currently available methods, challenges, and future research directions on SMS Spam filtering and detection techniques. Wang et al \cite{ wang2020comparative} conducted an empirical survey on word embeddings. Buchanan and Grant ~\cite{buchanan2001investigating} offered a brief description of Nigerian scam schemes, emphasizing that the growth of the Internet has aided the proliferation of cyber-crime. To classify Spam texts, Almeida et al. \cite{almeida2011contributions} used the SVM classifier. They used word frequency as a feature and discovered that SVM performed very well. Another study ~\cite{uysal2013impact} used SVM classifiers and k- Nearest Neighbor (kNN) to classify Spam text messages. The experimental results confirmed that using a combination of BoW features and structural features to classify Spam messages performed better. Some researchers have also proposed evolutionary methods for SMS Spam detection by assimilating the byte-level features of SMS \cite{abayomi2019review}. Researchers have recently started to use deep neural networks. To filter SMS Spam, Popovac et al. ~\cite{popovac2018convolutional} proposed a CNN-based architecture with one layer of convolution and pooling and achieved 98.4\% accuracy rate. Jain et al.~\cite{jain2019optimizing} used Long short-term memory (LSTM) model, for SMS Spam filtering. Their model attained a 99.01\% accuracy with the help of 6000 features and 200 LSTM nodes. They also employed various word embedding techniques including ConcepNet, WordNet and Word2Vec. \cite{barushka2018spam} proposed several deep neural network-based model and attained an accuracy of 98.51\%. 

Despite the preceding algorithms' success in identifying Spam messages, the existence of adversaries significantly degrades Spam filters' performance \cite{imam2019survey}. Graham-Cumming presented an approach against an individual user's Spam filter in a talk at the 2004 MIT Spam Conference~\cite{graham2004beat}. Random words were added into Spam mailings in this attack. A prior paper ~\cite{wittel2004attacking} attempted to broaden the scope of this attack by substituting regularly used English words with random phrases. After that, many adversarial attacks and countermeasures have been documented in a range of applications. Attacks on text classification jobs are typically carried out by altering characteristics or changing the text sequence's content \cite{li2018textbugger}. Several strategies have been identified in the field of adversarial attacks on Spam filters \cite{webb2005experimental,laskov2010machine}: injection of Ham words, obfuscation of Spam words, poisoning, alteration of labels, and synonym replacement. The impact of dictionary-based attacks and well-informed concentrated attacks, on the other hand, can be mitigated by using classifier weights ~\cite{peng2013revised}. 

In many ways, our research is different from that of our predecessors. No previous work, to our knowledge, has attempted to systematize and experimentally evaluate the SMS Spam literature from the perspective of security challenges. 

\section{Introducing The Super SMS Spam Corpus}\label{sec:data}
The Super SMS Spam corpus is a dataset of labeled SMS messages (reported over the last decade) that we collected for Spam research from the public and free for research sources. In any scientific study, having the most up-to-date, reliable, and representative data is critical. The most widely used SMS datasets including NUS SMS Corpus\footnote{https://github.com/kite1988/nus-sms-corpus} and UCI SMS Spam Collection\footnote{https://archive.ics.uci.edu/ml/datasets/SMS+Spam+Collection} used for Spam research are extremely outdated. Due to the exponential growth of online services and ongoing COVID19 epidemic, the threat landscape has significantly evolved in recent years, necessitating an updated dataset covering all the latest scams.

\subsection{Data Collection}
A comprehensive survey was conducted in order to identify, collect, and aggregate all of the public SMS datasets. Resultantly, we consolidated a corpus of 153,551 SMS instances from public and free-for-research sources. Details of different SMS datasets aggregated in the consolidated dataset are given in Table \ref{tab:consol}. 
Importantly, we manually crawled images of latest scam messages publicly shared on Twitter and those reported to the scam observatories including Scamwatch (Australian Competition and Consumer Commission's website for public education against scams) and Action Fraud (UK's national reporting center for fraud and cyber-crime)\footnote{\url{https://www.actionfraud.police.uk/}} to cover the recent landscape of SMS scams (categories and campaigns \cite{scamtypes}). This way, a list of 71 scam messages from Twitter and 141 Spam SMS messages from scam observatories were added to the corpus.

\subsection{Data Augmentation}\label{sec:transform}
All SMS messages in the consolidated dataset and newly gathered Spam messages from various volunteers and observatories were manually labelled using a set of carefully designed rules (list and details regarding derivation of rules are given in Appendix \ref{App:rules}). We eliminated discrepancies (duplicate and non-English messages) in the consolidated dataset and refined the data into a format suitable for further analysis by performing a series of processes. To this end, we first filter out SMS messages in non-English languages. For this purpose, 
in particular, we use a two pass filtering mechanism in order to filter the large amount of non-English SMS messages from the consolidated dataset. In the first pass, we use {\tt langdetect} (an open source Python library for language detection) \cite{langdetecturl} to determine the language of each SMS and filter out SMS in non-English languages. We then passed the filtered SMS messages returned by {\tt langdetect} to {\tt Googletrans} (Google API for language detection and translation) \cite{googlelangurl} to further filter out the non-English SMS messages. {\tt Googletrans} API was kept in the second phase of filtering due to the limitation of API calls for free user. 

Next, we removed the duplicate messages by importing the SMS corpus in {\tt mysql} table and deleted the duplicate rows. Lastly, we manually labelled all of the remaining unlabeled SMS messages (more than 47,022) as Spam and Ham (legitimate). Over all, we obtain a dataset of 52,814 unique English language SMSes from the consolidated dataset.

Lastly, the images crawled from Twitter and scam observatories were converted to text using online {\tt OCR} tool\footnote{https://www.onlineocr.net/}. 
Table~\ref{tab:super} presents ``Super Dataset'' which aggregates all SMSes from the augmented dataset, scam observatories, Twitter, and volunteers. 
The Super Dataset consists of 53,286 SMSes with 72.69\% and 27.31\% of Ham and Spam SMSes, respectively. 

\begin{table}[h]
\centering
\caption{Overview of SMS Spam datasets consolidated to generate an augmented dataset. 
}
\label{tab:consol}
\scalebox{0.80}{%
\renewcommand{\arraystretch}{1.2}
\tabcolsep=0.15cm
\begin{tabular}{r| r r r c} 
 \hline
 {\bf Dataset} & {\bf \# of SMSes} & {\bf Language} & {\bf Labeling} & {\bf Year}\\
 \hline\hline
UCI~\cite{almeida2013towards,almeida2011contributions} & 5,574 & English & Labelled & 2012
\\
 \hline
 NUS~\cite{chen2013creating} & 67,063 & English, Chinese & Unlabelled & 2015\\
 \hline
 Github1~\cite{githubbansalspamdetection} & 77,039 & English, Roman, Hindi & Unlabelled & 2019\\
 \hline
 Github2~\cite{githubnoorulhasansmsspam} & 557 & English & Labelled & 2018\\
 \hline
 Gupta ~\cite{gupta2018comparative} & 3,318 & English, Roman Hindi & Labelled & 2018\\ 
 \hline
 \hline
{Consolidated} & {153,551} & {Multi} & {Partial} & {-}\\
\textbf{{\makecell{Consolidated \\~[Augmented]}}} & \textbf{52,814} & \textbf{English} & \textbf{Partial} & \textbf{-}\\
 \hline
\end{tabular}}

\vspace{0.5cm} 

\caption{Characterisation of Super Dataset.
}
%
\label{tab:super}
\scalebox{0.80}{%
\renewcommand{\arraystretch}{1.2}
\tabcolsep=0.15cm
\begin{tabular}{r |c  c c} 
 \hline
 {\bf Dataset} & {\bf \# of SMSes} & {\bf Language} & {\bf Labeling} \\
 \hline\hline
  Consolidated [Augmented] (cf Table~\ref{tab:consol}) & 52,814 & English & Partial \\ 
 \hline
  DS7 [Volunteers] & 260 & English & Labelled   \\ 
 \hline
    DS8 [Scamwatch, ActionFraud] & 141 & English & Labelled \\ 
 \hline
    DS9 [Twitter] & 71 & English & Labelled \\ 
 \hline
 \hline
 \textbf{Super Dataset}  & \textbf{53,286} & \textbf{English} & \textbf{Labelled} \\
 \hline
\end{tabular}}
\end{table}
\section{Analysis Methodology}
\begin{figure}
\caption{Overview of experiment methodology.}\label{fig:protocol}
\centering
\includegraphics[width=0.49\textwidth]{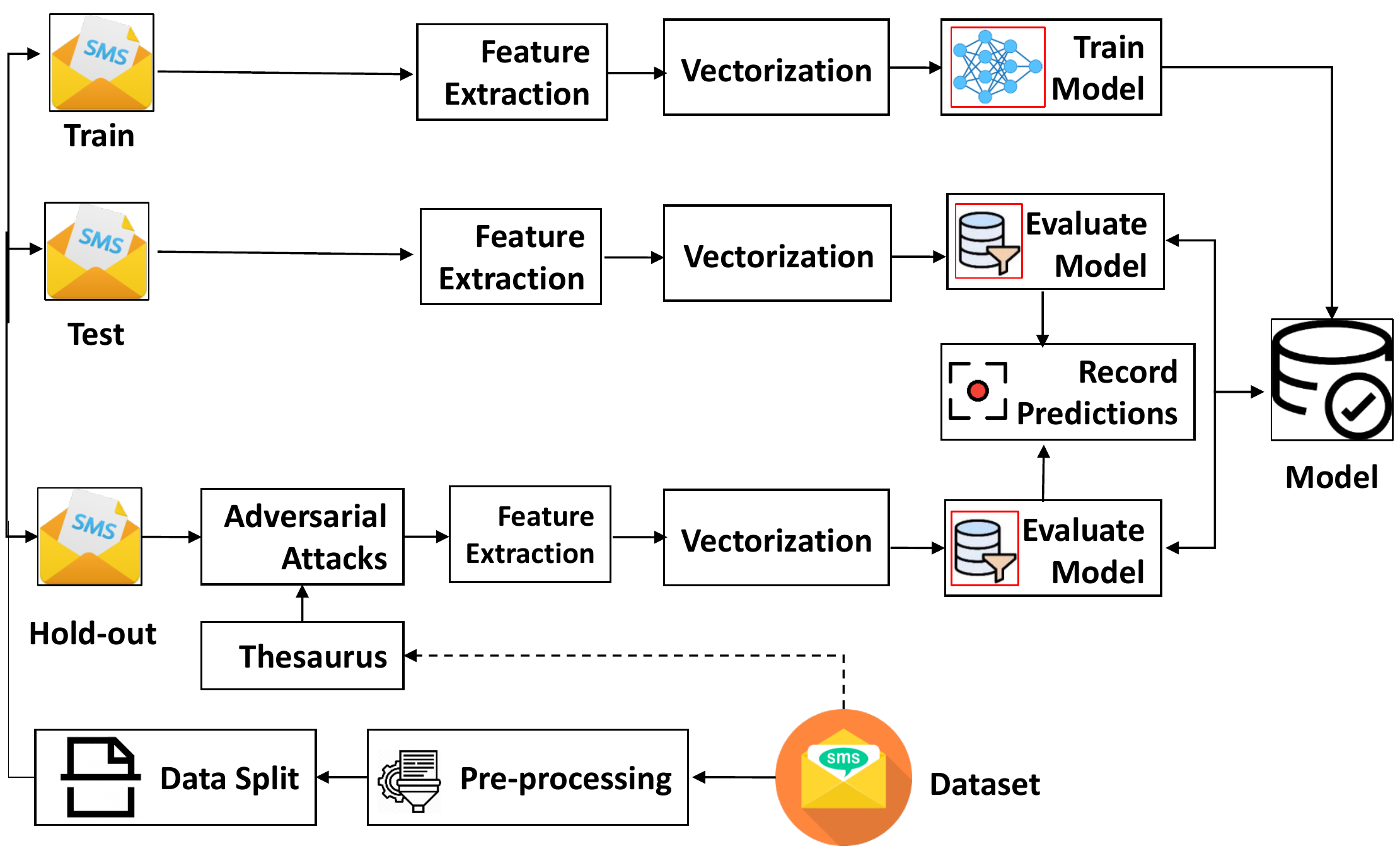}
\end{figure}

To identify the state-of-the-art in preventing SMS scams, we gathered existing techniques and counter measures from academic, industry, internet domain, and systematically categorize them. Figure~\ref{fig:protocol} depicts our experiment methodology to perform comparative analysis of various feature models and machine learning techniques.  

\subsection{Data Split}
Splitting the dataset is essential to build a reliable ML model and for an unbiased evaluation of classification performance. Common techniques for training and testing the data involve $k$-fold cross validation or splitting the original dataset into training (usually 80\%-70\%) and testing (usually 20\%-30\%) data. In this study, we divided the data set into three subsets: train (80\%), test (20\%), and hold-out. The hold-out split was created by randomly selecting 225 Spam SMS messages from all over the dataset. The train set was used to fit the ML model, whereas the test data set was used to estimate the performance of the model on data not used to train the model. The hold-out split was created for validation of Spam detection on diverse instances and adversarial evaluation (further elaborated in Section \ref{Adattack}). In order to split the data set into train and test splits, we utilised the {\tt train\_test\_split()} method of the {\tt scikit-learn} \cite{pedregosa2011scikit} library. Moreover, we removed the unnecessary variables such as punctuation, special symbols, and stop words from the dataset with the help of Python NLP library NLTK \cite{loper2002nltk}.

\subsection{Feature Extraction}\label{features}
\begin{figure}
\caption{A taxonomy of feature extraction/ representation techniques applied in this study.}\label{features_fig}
\centering
\includegraphics[width=0.49\textwidth]{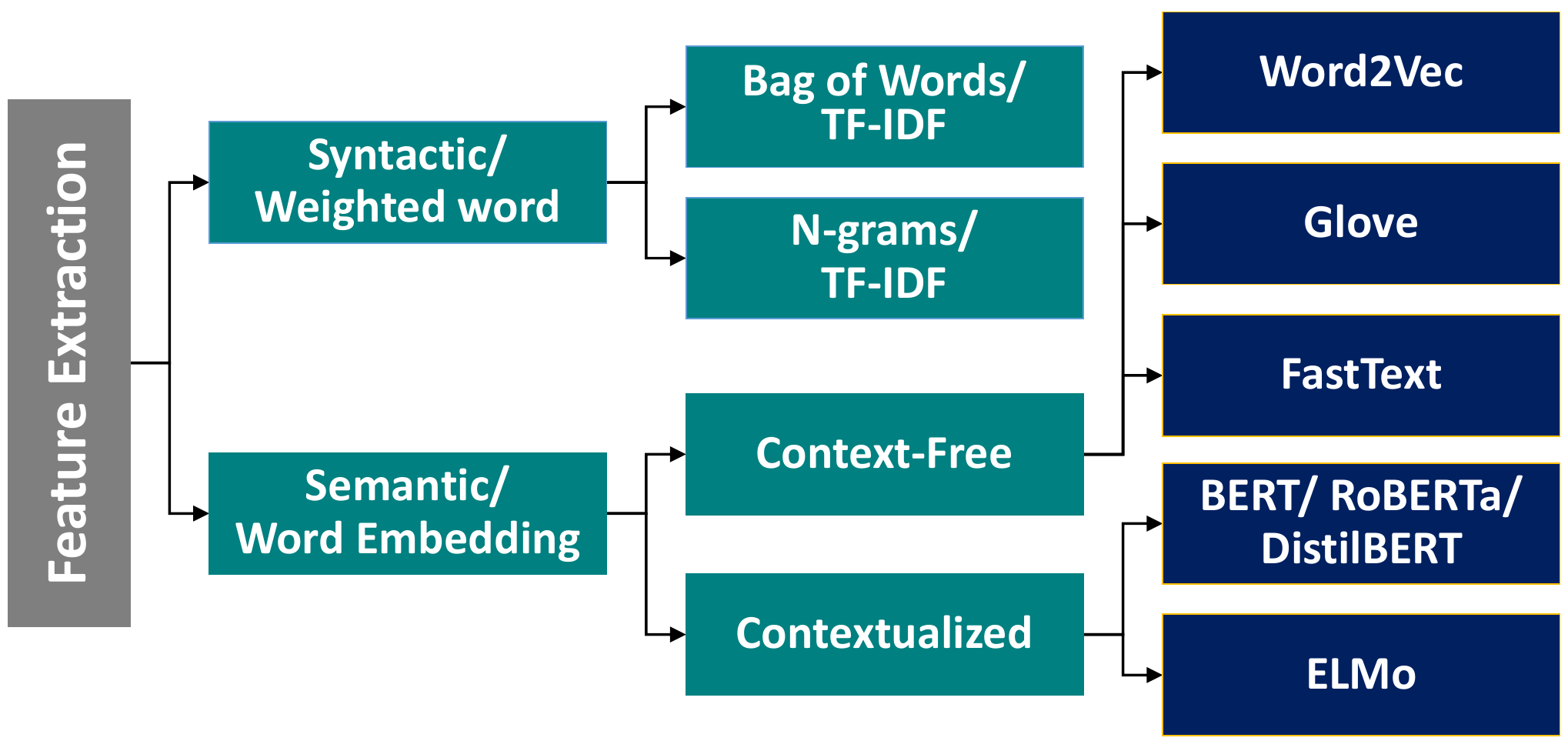}
\end{figure}

The unstructured text sequences in the SMS dataset need to be converted into a structured feature space before applying ML models. Therefore, we converted the list of words in each message into a feature vector. We generated both syntactic and semantic feature vectors of all the messages in the dataset to compare the impact of different feature types on the accuracy of classifiers. A taxonomy of text feature extraction techniques applied in this study is illustrated in Fig. \ref{features_fig}.

\subsubsection{Syntactic / Non-semantic Count based Vector space model}
We applied non-semantic techniques like bag of words (BoW) \cite{zhang2010understanding} and n-grams \cite{sidorov2012syntactic} with TF-IDF ~\cite{ko2012study} to convert raw text to numerical features. BoW captures word frequency in the SMS corpus, however, this model is sparse and doesn't consider word order. As a result, we employed N-grams, which are a document's sequence of N-words to count the word pairs (bigrams, trigrams etc.). The issue with N-gram is that even here, the resultant matrix is exceedingly sparse. Finally, we transformed the entire Bow and N-grams corpus into TF-IDF (term frequency-inverse document frequency) corpus. TF-IDF measures a word's importance in a document (SMS) and the corpus. Words that are uncommon in a document will have a high TF-IDF score. These mappings capture lexical features while ignoring semantics. We implemented this task with the help of the Python library scikit-learn.

\subsubsection{Semantic/ Word Embedding} 
Even while we have syntactic word representations, this does not imply that the model accurately represents the semantic meaning of the words. This constraint makes it difficult to interpret words within the model. To tackle this challenge, several academics focused on word embedding~\cite{mikolov2013efficient}. Word embedding is a feature learning approach in which each word or phrase in the vocabulary is mapped to an N-dimensional vector of real values in order to capture word semantics and context. Therefore, we created semantic feature vector for each of the following.

\subsubsection{Context-Independent Vector Space Model}\label{conind} 
To convert unigrams into intelligible input for ML systems, many word embedding approaches have been developed. In this section, we focused on the following popular classic word embeddings that have been and successfully implemented for text classification \cite{selva2021review}.

\textbf{Word2Vec} uses a neural network to learn word associations from large corpora and generate a vector representation for words or tokens and captures semantic similarity between words~\cite{mikolov2013exploiting}. Word2Vec creates a vector space with each unique word in the corpus, often with several hundred dimensions, such that words with similar contexts in the corpus are close to one another in the space. Moreover, the vectors for unknown words are randomly initialized using a generic normal distribution. 
In our study, we leverage {\tt Gensim} library~\cite{vrehuuvrek2011gensim} to implement both static and dynamic models of Word2Vec. 
With the static approach, the Word2Vec model uses pre-trained word embedding with 300-dimensional vectors. The vectors are kept static during training. Whereas, in the dynamic approach, the dimensional vectors are modified (fine tune the embeddings) during training phase. 
 
\textbf{GloVe~\cite{pennington2014glove}} In comparison to Word2Vec that utilises a window to establish local context, GloVe uses data from the whole text corpus to create an explicit word-context leveraging words' co-occurrence matrices. Each word in this method is represented by a high-dimensional vector and trained using the surrounding words throughout a large corpus. In order to achieve GloVe embeddings, we utilised the pre-trained 100 dimensional GLoVE model in Python, published by the Stanford NLP Group
\footnote{http://nlp.stanford.edu/data/glove.6B.zip}.

\textbf{fastText} is an extension of Word2Vec with improved efficiency and effectiveness~\cite{bojanowski2017enriching}. It addresses the out-of-vocabulary issues associated with Word2Vec feature model.  
%
In this work, we use Python's library {\tt fastText}\footnote{https://github.com/facebookresearch/fastText} for generating embeddings for our analysed machine learning models.

\subsubsection{Context-Dependent Vector Space Model} \label{cembed}  The context-independent vector space model provides context-free or static embeddings, which are incapable of capturing polysemy. Second, we only have one representation for a word, despite the fact that words have diverse features such as semantics, grammatical behavior, and register/connotations. To address the aforementioned issues, the "Contextualized Word Representations" technique was introduced, which aims to capture word semantics in various settings in order to solve the problem of polysemous words and their context-dependence \cite{wang2020comparative}. Here, we focused on the following contextualised word embeddings.

\textbf{Bidirectional Encoder Representations from Transformers (BERT)~\cite{devlin2018bert}} BERT is a powerful transformer-based architecture that has demonstrated state-of-the-art performance on different NLP tasks \cite{vaswani2017attention}. Unlike embedding representations of individual words (such as Word2Vec, GloVe, or fastText), BERT is backed by Transformer and it’s core principle - attention, which understands the contextual relationship between different words. In this study, we generated the BERT embeddings by implementing a pre-trained BERT model using the SimpleTransformers \cite{thilina22} package in Python (further details in Appendix \ref{app:explan}).

\textbf{Embedding from Language Model (ELMo)~\cite{ELMo}} ELMo is a bidirectional Language Model (biLM) that extracts multi-layered word embeddings. ELMo word vectors address Polysemy, where a word has several meanings. These word vectors are pre-trained, deep bidirectional language model (biLM) functions. They can be easily integrated into current models and progress complex NLP tasks like question answering, textual entailment, and sentiment analysis. In our implementation, we accessed ELMo via TensorFlow Hub (a library / repository of trained ML models that enables transfer learning)\footnote{https://www.tensorflow.org/hub} using Python.

\subsection{Classification Techniques}
\begin{figure}
\centering
\includegraphics[width=0.5\textwidth]{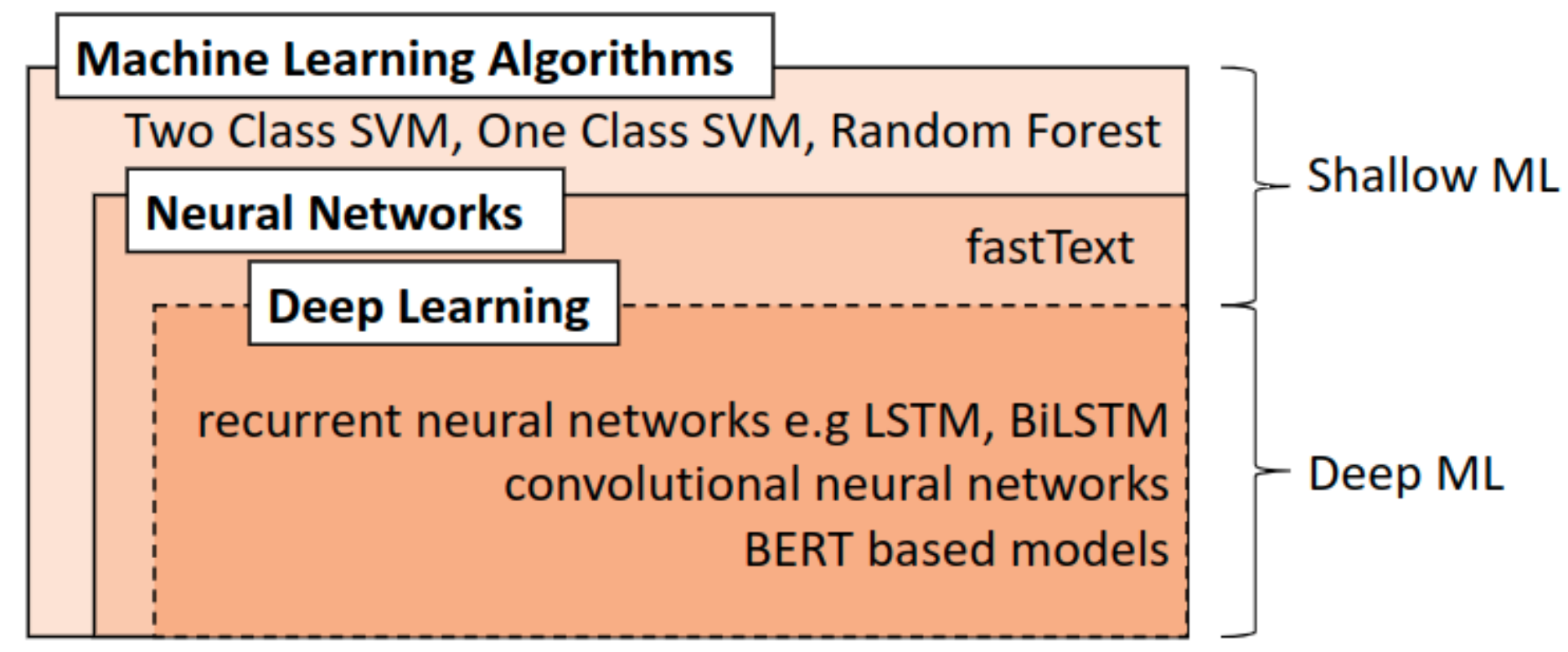}
\caption{Venn Diagram of machine learning classifiers applied in this study (inspired by Goodfellow et al. ~\cite{goodfellow2016deep}). 
} \label{venn}
\end{figure}

Choosing the right classifier/ architecture for model building/ training is one of the most critical steps in the Spam detection pipeline. However, researchers have difficulties in defining acceptable structures, architectures, and methodologies for text classification. In this section, a brief overview of different classifiers and architecture implemented in this study are discussed. Figure \ref{venn} depicts various machine learning based approaches analysed in this work. 


Given that there is no solid theory on how to map algorithms onto problem types; instead, a practitioner should research the literature work for the specific problem and conduct controlled trials to determine which algorithm and algorithm configuration performs best for a certain classification job.
At this stage, we designed a first set of experiments to evaluate how well the most popular techniques of text classification would perform on the original messages. For this purpose, we developed several models using different ML approaches. The approaches considered in this study are stated below.
\subsubsection{Two-class Classification}
We started with the conventional methods of text classification such as two-class classification or binary classification ~\cite{kowsari2019text}. Some algorithms are primarily built for binary classification and do not natively handle more than two classes; for example, Support Vector Machines (SVM) \cite{vapnick1998statistical}. SVM have shown to be one of the most powerful and efficient state-of-the-art classification algorithms for handling the Spam problem \cite{torabi2015efficient}. They're supervised learning models that examine data and find patterns that may be used to categorize and investigate the relationship between variables of interest. SVM outperforms many classifiers on Spam detection by employing content-based filtering to find particular features in the text \cite{scholkopf2002learning}. In the two-class classification approach designed in this study, we implemented a binary SVM using Scikit-learn.
\subsubsection{One Class Classification}
Next, we covered the one-class classification (OCC) ~\cite{khan2014one} approach which includes methods for detecting outliers and anomalies. The one-class algorithms are recognition-based, with the goal of recognizing data from one class and rejecting data from all others.

In this approach, we employed One Class Support Vector Machine (OCSVM) ~\cite{manevitz2001one}, which is a natural extension of SVMs. The OCSVM is an unsupervised technique for learning a decision function for novelty detection, which involves categorising fresh data as similar or dissimilar to the training set.
Generally, an OCSVM or one class learning is suitable for imbalanced dataset (similar to our dataset) because only the majority class is needed to train a given ML model. Moreover, such an approach is useful when it is hard to create outliers. In many practical scenarios, it is hard and expensive to create anomalies, certainly the case for fraud detection which happens rarely. Also, if you have a new type of fraud that has never been encountered before, it would be easy for OCSVM to detect it, whereas a traditional classification model may struggle to do so ~\cite{hodge2004survey}. 

\subsubsection{Positive and Unlabeled (PU) Learning}
After the supervised (two class SVM) and unsupervised (OCSVM) approaches, we next considered positive and unlabeled data (PU) learning \cite{hu2019learning}, which is a semi-supervised approach that learns from the positive cases in the data and uses them to predict the unknown cases. There are two primary methods for implementing PU learning. These include PU bagging and a two-step process. The PU bagging is basically an ensemble of ensembles; training many ensemble classifiers in parallel. Each ensemble balances the classes according to the size of the positive class. On the other hand, the two-step method is a more complicated way to learn about PU. It uses ML techniques to change the labels of data while training and takes longer time to train. Therefore, we decided to opt the PU bagging approach (quicker than normal ensemble methods) and implemented it with random forest (RF)~\cite{debarr2009spam} classifier. 

\subsubsection{Neural Networks (NNs)} Most of the state-of-the-art text classification is based on neural networks \cite{kowsari2019text}. As discussed earlier in Sec \ref{conind}, Facebook recently released a lightweight shallow neural network library (fastText) for unsupervised (embedding) representations of words and supervised text classification ~\cite{joulin2016bag}. fastText employs a hierarchical classifier that decreases the training and testing time complexity of text classifiers from linear to logarithmic in terms of the number of classes. It also exploits the fact that classes are imbalanced (some classes appearing more often than other, similar to our Spam dataset) by using the Huffman algorithm \cite{van1976construction} leading to further computational efficiency.

\subsubsection{Deep Learning}\label{dlclassifiers}
Lately, various DL architectures have emerged and outperformed shallow ML in NLP, image classification, etc \cite{lecun2015deep}. Unlike shallow ML, DL can create models directly on high-dimensional raw data using automated feature learning. Therefore, we finally considered several state of the art DL models that have been successfully tested for text or Spam classification by researchers. Depending on the learning task, DL offers various architectural variants such as recurrent neural networks (RNNs) \cite{mandic2001recurrent}, convolutional neural networks (CNNs) \cite{lai2015recurrent}, transformer \cite{tay2020efficient} etc. Architectural variations are distinguished primarily by the layers, neuronal units, and connections employed. Below, we describe each DL model used for Spam detection in this study. 

\textbf{Recurrent Neural Network (RNN)}:
RNN \cite{mandic2001recurrent} is one of the most utilized and successful DL architectures for text classification. RNNs are typically used to classify text using LSTM or GRU \cite{kowsari2019text}. In this work, we implemented the three widely used RNN based models such as LSTM \cite{hochreiter1997long}, BiLSTM \cite{graves2005framewise} and BiGRU \cite{qing2019novel} (brief introduction of each in Appendix \ref{app:explan}) for Spam detection using Keras \cite{gulli2017deep} and TensorFlow2 \cite{abadi2016tensorflow} in Python.

\textbf{Convolutional Neural Network (CNN)} Despite several advantages, RNN-based designs might be biased when later words are more significant. CNN models were developed to address this bias by utilising a max-pooling layer to identify \emph{discriminative} words in text input \cite{lai2015recurrent}. Moreover, CNNs can help us with parallelization, local dependencies and distance between positions. Consequently, we implemented a 1D CNN for the Spam detection since it has demonstrated to perform effectively on text categorization with comparatively small parameter tuning ~\cite{chen2015convolutional}. Moreover, we employed a generic temporal convolutional network (TCN) \cite{bai2018empirical}, which is a dilated-causal variant of the CNN and a viable alternative to recurrent architectures since it is capable of dealing with extended input sequences without experiencing vanishing gradient problems. Additionally, we constructed a model based on ensemble learning by merging a 1D CNN with a single Bidirectional GRU (BiGRU). One of the reason for ensembling with BiGRU is that it performs effectively with temporal data because it incorporates both prior and subsequent information in the sequence. In this study, we implemented all of the convolutional architectures using Keras and TensorFlow2 in Python.

\textbf{Transformers} CNNs are successful with shorter phrases, but gathering word associations in longer sentences is tedious and impractical. In 2017, Google introduced Transformers to deal with the long-range dependency challenge \cite{vaswani2017attention}. Transformer avoids recursion via the use of attention mechanisms and positional embeddings to comprehend phrases in their entirety. In 2018, BERT was open-sourced by Google \ref{cembed}). With just one extra output layer, the pre-trained BERT model may be fine-tuned to build state-of-the-art classifiers for a variety of tasks, including text classification, without requiring significant task-specific architectural adjustments \cite{devlin2018bert}. Recently, modified BERT models have been proposed by researcher to enhance BERT's prediction metrics or computational performance, and these models have been demonstrated to attain state-of-the-art results to counter email phishing~\cite{lee2020catbert} and applied to filter multilingual Spam messages \cite{cao2020bilingual} with promising results. In this study, in addition to the original BERT, we employed some of the modified models proposed by researchers that outperform BERT in many tasks. These models include DistilBERT \cite{sanh2019distilbert}, RoBERTa \cite{liu2019roberta}, and Cross-lingual Language Model (XLM) \cite{conneau2019cross} (brief introduction of each is given in Appendix \ref{app:explan}). We implemented all of the BERT-based models using the SimpleTransformers package in Python.

\subsubsection{Adversarial Attacks} \label{Adattack}
Effective adversarial attacks depend on the attacker's knowledge of the targeted ML model. Attacks can be classed as black-box or white-box based on target model information and targeted or non-targeted based on whether the erroneous output is desired \cite{wang2019towards}. In a white-box scenario, an attacker can access target models and training data and generate perfect adversarial instances. Black-box attacks are used in situations when the attacker has limited knowledge of the target model \cite{lei2018discrete}. An attacker must repeatedly test the model with different inputs to find a flaw. Adversarial attacks in the text domain are further classified as char-level, word-level, sentence-level, and multi-level based on the perturbation units used to generate adversarial samples to produce hostile samples that trick detectors. Multi-level attacks combine many perturbation attacks to produce an unnoticeable and high-successful attack.
\begin{figure}[hbt!]
\caption{A taxonomy of eight different adversarial attacks conducted in this study. The analysed attacks are grouped into perturbations onto character, words, sentence, and miscellenous, are shown in the figure.}\label{attacks}
\centering
\includegraphics[width=0.48\textwidth]{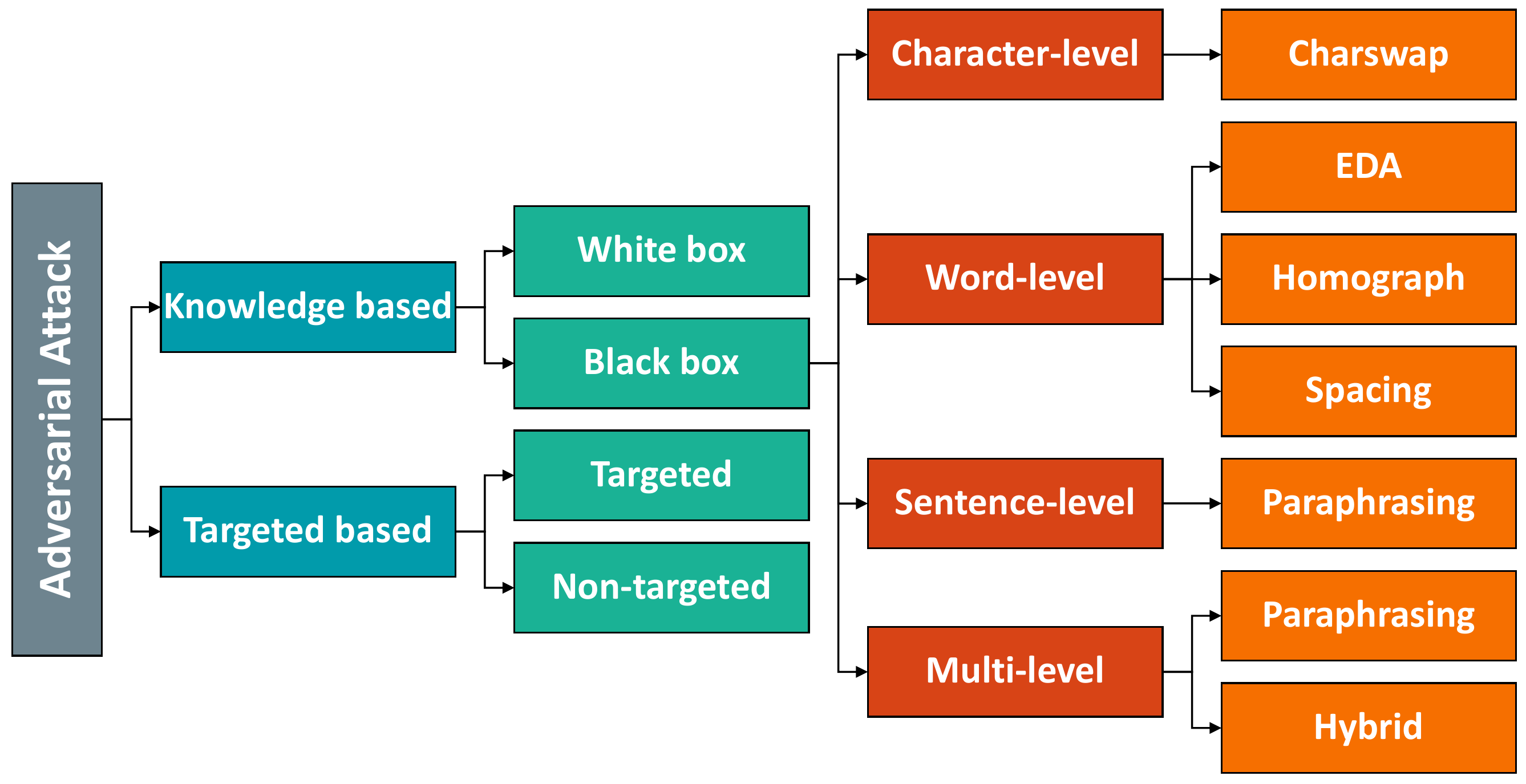}
\end{figure}

In order to assess the robustness of ML models, we reviewed the literature on evasion tactics in text categorisation and Spam detection. Currently, Spam detection models are examined using conventional blackbox adversarial mechanisms \cite{rojas2021using}. However, due to the convergence of mobile networks and the internet, scammers may likely adapt the techniques used for tricking internet users to deceive mobile users as they go online. For instance, Smishing is an SMS Spam campaign in which scammers send an enticing text message to deceive receivers into opening a link and transferring the victim private information or downloading malware \cite{ulfath2022detecting}. Consequently, it is necessary to proactively seek such approaches.

Punycode attack is a homograph technique used in URL phishing \cite{gabrilovich2002homograph} \cite{ilca2021phishing}. The attacker creates a homograph URL by choosing similar letters from other languages, especially Cyrillic or non-Latin. To the human eye, these Cyrillic glyphs or other similar non-Latin characters can easily be confused with their Latin counterparts. Computers, however, interpret it differently, since different hex codes are assigned to them. This technique may not only be used in Smishing but other Spam SMS campaigns by encoding the top Spam keywords in order to bypass the Spam filters.  The homograph technique like Punycode haven't been explored for SMS Spam problem earlier. Figure~\ref{attacks} presents taxonomy of adversarial attacks performed in this study. We created the adversarial texts with the following principle in mind: they must retain the same semantic meaning as the originals. Moreover, we developed a thesaurus of the 200 most frequent keywords from the Spam messages in the dataset with the help of the Python library WordCloud \cite{heimerl2014word}. 

In the following, we elaborate on the adversarial attacks employed in this study: 
\begin{itemize}
    \item \textbf{Paraphrasing} A popular attack for fooling Spam filters by replacing vocables with synonyms or similar phrases, rendering the message unrecognizable to the detection model. We carried out this attack in 2 folds. In the first fold, we used a commercial tool Quillbot\footnote{https://quillbot.com/} to restructure the sentences. In the second fold, we replaced all of the Spam keywords from thesaurus with its synonyms in each SMS message. 
    \item \textbf{Easy Data Augmentation (EDA)} This tactic bypasses Spam filters by randomly removing, swapping, replacing, or adding a word's synonym in a sentence. We carried out this attack with the help of TextAttack \cite{morris2020textattack}.
    \item \textbf{Homograph} This attack involves replacing thesaurus's keywords found in SMS with its punycode, by generating homoglyphs for each character in the corresponding Spam word. We generated the homoglyphs with the help of an online homoglyph attack generator \footnote{https://www.irongeek.com/homoglyph-attack-generator.php} tool.
    \item \textbf{Spacing} In this technique, spaces are added between the characters of Spam keywords to evade detection. We designed this attack manually by adding spaces between the characters of the thesaurus's keywords found in SMS.
    \item \textbf{Charswap} This type of attack randomly substitute, delete, insert, and swap adjacent characters of Spam keywords in the message. We utilise the TextAttack framework to generate adversarial examples of this type.
    \item \textbf{Hybrid} This type of attack involves combination of different attacks. We designed this attack using a combination of spacing technique and replacement of visually same characters in Spam words.
\end{itemize}
\subsection{Evaluation Metrics}
For the evaluation of ML models, multiple aspects have to be taken into account, such as performance, computational resources, interpretability and robustness. Performance-based metrics quantify how effectively a model achieves the learning task's purpose. In this work, we evaluated the models using the metrics of Precision (PR), Recall (RE), Accuracy (ACC) and F1-score (F1) \cite{forman2003extensive}. The latter is considered the most appropriate metric for this particular task, considering that the dataset is highly unbalanced. Formally, these metrics are defined by the following equations:
\begin{equation}
    Precision = \frac{TP}{TP+FP}
\end{equation}
\begin{equation}
    Recall = \frac{TP}{TP+FN}
\end{equation}
\begin{equation}
    Accuracy = \frac{TP+TN}{TP+TN+FP+FN}
\end{equation}
\begin{equation}
    F1 = \frac{2*Precision*Recall}{Precision+Recall} = \frac{2*TP}{2*TP+FP+FN}
\end{equation}

Here, we consider Spam SMSes as positives while non-Spam (i.e., Ham) ones as negatives. We represent the total number of Spams and Ham SMSes by $P$ and $N$, respectively. Whereas, in the above equations, the TP represents the number of true positives (i.e., correctly classified as Spam) and FP shows the number of false positives (i.e., wrongly classified as Spam). Similarly, we denote true negatives and false negatives by, TN and FN, respectively. TN shows the number of SMSes correctly classified as Ham while FN represents the number of SMSes that are wrongly classified Ham.

\section{Comparative Evaluation}
In this section, we evaluated the efficacy/competence of the models in two steps. The first set of experiments approximated the performance of the models in order to evaluate their efficacy in classifying Spam and legitimate messages, while the second set of experiments assessed the robustness of the models. 

\subsection{Results Using The Traditional ML Approach}
First, the ML models were evaluated without adversarial examples. Initially, we encoded the messages in the Train and Test splits with different representation models (ref para \ref{features}) to obtain feature vectors which were then fed to the classification algorithms. Each classifier was trained with the encoded vectors of the Train split along with their respective labels; once trained, their performance was evaluated on the Test split.

We started the experimentation with traditional two-class SVM (TCSVM). We trained several models of TCSVM with different syntactic and semantic features including BoW, n-grams, Word2Vec and GloVe, extracted from the train split of the super dataset. We then used the trained models to classify the instances of the test set. Next, we fed the same set of features to the OCSVM and PU classifiers. However, for OCSVM and PU, we use only Spam messages of the train split since these two classifiers only require samples from the positive class. The trained models are then evaluated with the messages of the test split for training. The performance evaluation results of the aforementioned ML models are summarized in Table \ref{tab:perform-ml}. The performance metrics are reported and grouped by the encoding/ feature model and classification algorithms. 

Roughly speaking, the results varied widely depending on the classification algorithm and feature model. With all the feature models including TFIDF, Word2Vec and GloVe, the OCSVM performance remains significantly lower than other models, both in terms of f1-score and accuracy. On the other side, good performance was demonstrated by most of the TCSVM, PU and fastText models, however, TCSVM with Word2Vec produced the highest among all the shallow ML models with 83\% and 95.5\% F1-score and accuracy respectively. It is pertinent to mention to that all of the classifiers demonstrated good performance with BoW. With the exception of OCSVM, the models perform better with classic word embedding techniques like Word2Vec and GloVe, producing higher f1-score in comparison to other models. The PU and TCSVM with Word2Vec and GloVe embeddings outperforms all other models including OCSVM and fastText. Lastly, the models with n-grams in this experiment cannot do much despite trying different “n” grams and having so many hyper parameter tuning.
\subsection{Results Using The DL Approach}
Next, we evaluated the state-of-the-art DL models (discussed in section \ref{dlclassifiers}), proposed by academia/researchers in top-tier conference proceedings, workshops, symposiums, journals, and pre-prints published on arXiv\footnote{https://arxiv.org/}. Like in shallow ML evaluation, the DL models were first trained on the training split of the super dataset with different embeddings/ representation models and then evaluated on the test split. Some of the DL classifiers are also trained with random word embedding(WE-Random). For WE-Random, we simply define a randomly initialised trainable embedding layer. Thus, it is not pre-trained but learned from scratch with the rest of the downstream model parameters and corrected during training. Table \ref{tab:perform-dl} shows performance evaluation of DL models. 

In comparison to traditional ML models, the variability in the accuracy of the DL models is less noticeable. All of the classifiers obtained an accuracy greater than 90\%, with the Ensemble model of CNN-BiGRU with Word2Vec(static) achieving the highest accuracy (98.4\%), slightly higher than LSTM and BiLSTM (98.3\% \& 98.2\% respectively). On contrary, the variability in the f1-score of the models is quite high. The RoBERTa models produced the lowest f1-score whereas LSTM, BiLSTM, CNN, and TCN achieved a high score of 97\% each.

From the high accuracy of shallow ML and deep ML models in Table \ref{tab:perform-ml} and \ref{tab:perform-dl}, it is evident that all of the models produce impressive results for the Ham (legitimate messages) class, whereas for the Spam class the results are very poor in most of the models, as shown by the f1-score. Out of a total of sixteen shallow ML models, only two models (TCSVM with Word2Vec and GloVe) achieved greater than or equal to 80\% f1-score (83\% being the highest achieved by TCSVM with Word2Ve). Similarly, only four deep ML models (LSTM, BiLSTM, CNN with Random and GloVe embeddings) out of fifteen models crossed the 90\% mark.
\begin{table}[hbt!]
\renewcommand{\arraystretch}{1.15}
\tabcolsep=0.15cm
\begin{center}
\caption{Performance evaluation of {\it shallow} ML classifiers with the Super dataset (cf. Table~\ref{tab:super}). Here PR, RE, FS, ACC, and CM represent precision, recall, F1-score, accuracy, and confusion matrix, respectively.}\label{tab:perform-ml}
\newcolumntype{a}{>{\columncolor{red!50}}c}
\newcolumntype{b}{>{\columncolor{green!50}}c}
\begin{tabular}{ c | c | b | >{\columncolor{yellow!50}} c | a | a | a}
 \hline
 \textbf{Feature}& & \multicolumn{5}{a}{\bf Performance Metrics}\\
\cline{3-7}
 \textbf{Model} & \textbf{Classifier} & \textbf{PR}&\textbf{RE} &\textbf{F1} &\textbf{ACC} &\textbf{CM}$\big(\begin{smallmatrix} TN & FP\\ FN & TP \end{smallmatrix}\big)$ \\ [0.5ex]
 \hline\hline
  \multirow{ 3}{*}{\makecell{BoW \\ (TF-IDF)}} & TCSVM & 90\% & 59\% & 71\% & 93.3\% & $\big(\begin{smallmatrix} 9917 & 104\\ 682 & 983 \end{smallmatrix}\big)$ \\
  & OCSVM & 53\% & 40\% & 45\% & 86.3\% & $\big(\begin{smallmatrix} 9427 & 593\\ 1007 & 658 \end{smallmatrix}\big)$\\
  & PU & 87\% & 71\% & 78\% & 94.4\% & $\big(\begin{smallmatrix} 9839 & 182\\ 476 & 1189 \end{smallmatrix}\big)$\\
  \hline
  \multirow{ 3}{*}{\makecell{Bigram \\ (TF-IDF)}} & TCSVM & 94\% & 24\% & 38\% & 89.0\% & $\big(\begin{smallmatrix} 9997 & 24\\ 1263 & 402 \end{smallmatrix}\big)$\\
  & OCSVM & 7\% & 35\% & 12\% & 27.1\% & $\big(\begin{smallmatrix} 2587 & 7433\\ 1086 & 579 \end{smallmatrix}\big)$\\
  & PU & 90\% & 30\% & 45\% & 89.6\% & $\big(\begin{smallmatrix} 9966 & 55\\ 1162 & 503 \end{smallmatrix}\big)$\\
  \hline
  \multirow{ 3}{*}{\makecell{Trigram \\ (TF-IDF)}} & TCSVM & 94\% & 7\% & 12\% & 86.6\% & $\big(\begin{smallmatrix} 10014 & 7\\ 1554 & 111 \end{smallmatrix}\big)$\\
  & OCSVM & 12\% & 80\% & 21\% & 12.4\% & $\big(\begin{smallmatrix} 121 & 9899\\ 332 & 1333 \end{smallmatrix}\big)$\\
  & PU & 99\% & 9\% & 16\% & 87.0\% & $\big(\begin{smallmatrix} 10019 & 2\\ 1521 & 144\end{smallmatrix}\big)$\\
  \hline
  \multirow{ 3}{*}{Word2Vec} & TCSVM & \cellcolor{red!100}89\% & \cellcolor{red!100}78\% & \cellcolor{red!100}83\% & \cellcolor{red!100}95.5\% &  \cellcolor{red!100}$\big(\begin{smallmatrix} 9794 & 156\\ 367 & 1298 \end{smallmatrix}\big)$\\
  & OCSVM & 17\% & 94\% & 29\% & 34.4\% & $\big(\begin{smallmatrix} 2385 & 7395\\ 108 & 1556 \end{smallmatrix}\big)$\\
  & PU & 74\% & 84\% & 79\% & 93.5\% & $\big(\begin{smallmatrix} 9458 & 492\\ 259 & 1406 \end{smallmatrix}\big)$\\
  \hline
  \multirow{ 3}{*}{GloVe} & TCSVM & 86\% & 77\% & 81\% & 94.9\% & $\big(\begin{smallmatrix} 9749 & 206\\ 390 & 1275 \end{smallmatrix}\big)$\\
  & OCSVM & - & - & - & 27.8\% & $\big(\begin{smallmatrix} 3289 & 8521\\ 0 & 0 \end{smallmatrix}\big)$\\
  & PU & 67\% & 82\% & 73\% & 92.4\% & $\big(\begin{smallmatrix} 9359 & 596\\ 269 & 1196 \end{smallmatrix}\big)$\\
  \hline
  fastText & fastText & 92\% & 54\% & 68\% & 92.8\% & $\big(\begin{smallmatrix} 9940 & 82\\ 765 & 900 \end{smallmatrix}\big)$\\
  \hline
  \hline
\end{tabular}
\end{center}
\end{table}

\begin{table}[hbt!]
\begin{center}
\caption{Performance evaluation of Deep ML classifiers with Super Dataset (cf. Table \ref{tab:super}).}\label{tab:perform-dl}
\newcolumntype{a}{>{\columncolor{red!50}}c}
\newcolumntype{b}{>{\columncolor{green!50}}c}
\tabcolsep=0.1cm
\renewcommand{\arraystretch}{1.2}
\scalebox{0.97}{\begin{tabular}{ r | r | b | >{\columncolor{yellow!50}} c | a }
 \hline
 \textbf{Classifier} & \textbf{Embedding} &\textbf{F1} & {\bf ACC} & \textbf{CM}$\big(\begin{smallmatrix} TN & FP\\ FN & TP \end{smallmatrix}\big)$ \\ [0.25ex]
 \hline\hline
{BERT} & bert-base-uncased & {79\%} & 94.9\% & $\big(\begin{smallmatrix} 9974 & 48\\ 552 & 1113 \end{smallmatrix}\big)$ \\
 \hline
{ELMo} & ELMO & {80\%} & 95\% & $\big(\begin{smallmatrix} 9977 & 44\\ 537 & 1128 \end{smallmatrix}\big)$\\
\hline
{RoBERTa} & roberta-base & {53\%} & 90.9\% & $\big(\begin{smallmatrix} 10014 & 8\\ 1056 & 609 \end{smallmatrix}\big)$\\
\hline
{XLM-RoBERTa} & xlm-roberta-base & {52\%} & 90.5\% & $\big(\begin{smallmatrix} 9974 & 48\\ 1063 & 602 \end{smallmatrix}\big)$\\
\hline
{DistilBERT} & distilbert-base-uncased & {78\%} & 94.8 & $\big(\begin{smallmatrix} 9996 & 26\\ 582 & 1083 \end{smallmatrix}\big)$\\
\hline
LSTM & WE-Random & {97\%} & 98.3\% & $\big(\begin{smallmatrix} 10086 & 78\\ 164 & 3594 \end{smallmatrix}\big)$\\
\hline
BiLSTM & WE-Random & {97\%} & 98.2\% & $\big(\begin{smallmatrix} 10059  & 105\\ 152 & 3606 \end{smallmatrix}\big)$\\
\hline
\multirow{2}{*}{CNN} & WE-Random & {97\%} & 98.3\% & $\big(\begin{smallmatrix} 10044 & 120\\ 119 & 3639 \end{smallmatrix}\big)$\\
& {GloVe (static)} & {97\%} & 98.2\% & $\big(\begin{smallmatrix} 10039 & 125\\ 129 & 3629 \end{smallmatrix}\big)$\\
\hline
\multirow{3}{*}{TCN} & WE(Random) & {80\%} & 97.6\% & $\big(\begin{smallmatrix} 12924 & 165\\ 172 & 661 \end{smallmatrix}\big)$\\
& {Word2Vec (static)} & {76\%} & 97.3\% & $\big(\begin{smallmatrix} 12961 & 128\\ 244 & 589 \end{smallmatrix}\big)$ \\
& {Word2Vec (dynamic)} & {69\%} & 94.1\% & $\big(\begin{smallmatrix} 12180 & 76\\ 742 & 923 \end{smallmatrix}\big)$ \\
\hline
\multirow{3}{*}{\makecell{Ensemble \\ (CNN-BiGRU)}} & WE-Random & {80\%} & 95.8\% & $\big(\begin{smallmatrix} 12190 & 66\\ 518 &  1147 \end{smallmatrix}\big)$ \\
& {Word2Vec (static)} & \cellcolor{red!100}86\% & \cellcolor{red!100}98.4\% & \cellcolor{red!100}$\big(\begin{smallmatrix} 12993 & 96\\ 133 & 700 \end{smallmatrix}\big)$ \\
& {Word2Vec (dynamic)} & {80\%} & 95.7\% & $\big(\begin{smallmatrix} 12175 & 81\\ 511 & 1154 \end{smallmatrix}\big)$ \\
\hline
  \bottomrule
\end{tabular}}
\end{center}
\end{table}

\section{Adversarial Evaluation: Robustness Assessment of ML Models}\label{sec:aml}
The second set of experiments focused on quantitatively evaluating the influence of adversarial examples on our trained models. In order to assess the robustness of our trained models, we employed six different sets of black-box attacks discussed previously (ref para \ref{Adattack}).
To do this, we used the Hold-out split, together with the thesaurus, to carry out the attacks. Each attack attempted to substitute all Spam words in the message that were found/matched with the thesaurus (see examples in Table \ref{tab:ad-examples}). Once the attack was completed, the modified messages were encoded with different representation models to obtain the corresponding feature vectors, and then fed to the previously trained models in order to evaluate their performance with the same metrics. To begin, we first passed the actual instances of the Holdout split to the trained models in order to estimate their performance on positive samples and get a baseline result. Following that, the adversarial examples were passed to the model to assess its robustness. The results of the adversarial evaluation of shallow ML and deep ML models are presented in Table \ref{tab:robust-ml} and Table \ref{tab:robust-dl}, respectively.
\begin{table}
\begin{center}
\caption{Instances of Adversarial Attacks (Spam messages) colored \textcolor{red}{red}.}
\label{tab:ad-examples}
\def\arraystretch{0.3}%
\scalebox{0.95}{\begin{tabular}{p{1.7cm}|p{6cm}} 
 \hline
 {\bf \textcolor{red}{Attack}} & {\bf SMS Text} \\
 \\
 \hline
Actual SMS & You may get a \$750 Economic Support Payment. For more details, Click https://xxx.info/covid/ \\
 \hline
\textcolor{red}{Paraphrasing} & A \$750 Economic Support Compensation may be available to you. For further details https://xxx.info/covid/ \\ 
 \hline
 \textcolor{red}{Charswap} & You Qmay gjt a \$7Q0 Economic Support Payment. For fruther detials https://XXXXX.info/COVID/?r=xxxx \\ 
 \hline
 \textcolor{red}{EDA} & You may get a \$750 Economic Support Payment https://XXXXX.info/COVID/?r=xxxx \\ 
 \hline
 \textcolor{red}{Homograph} & You may get a \$750 Economic Support $payment$ https://XXXXX.info/COVID/?r=xxxx\\
 \hline
 \textcolor{red}{Spacing} & You may get a \$750 Economic Support p a y m e n t. For more details, C l i c k https://xxx.info/covid/ \\
 \hline
 \textcolor{red}{Hybrid} & You may get a \$750 Economic Support p @ y m e n t. For more details, C 1 i c k https://xxx.info/covid/\\
 \hline
 \hline
\end{tabular}}
\end{center}
\end{table}

For shallow ML models, the spacing tactic remains the most effective technique, evading eight models the most. The spacing tactic was most effective in evading all classifiers trained using BoW/TF-IDF (TCSVM, OCSVM, PU), GloVe (TCSVM, OCSVM, PU), and fastText as well as the TCSVM-Word2Vec model, which demonstrated the highest performance in the previous section. Charswap, on the other hand, was the second most successful attack against the six models presented in Table \ref{tab:robust-ml}, completely evading three (Trigram/TCSVM, Trigram/PU, Word2Vec/PU), while the Bigram/TCSVM and Bigram/PU models survived to achieve a little score. Moreover, the Homograph/Punycode attacks also proved to be quite efficient against all the models except those trained with BoW.

In comparison to shallow ML models, the deep ML models showed some robustness to different adversarial techniques. However, their performance degraded significantly in the presence of the spacing technique. The spacing evasion technique proved to be the most successful attack against thirteen models out of the total fifteen models presented in Table \ref{tab:robust-dl}. The hybrid attack proved to be the second most successful attack, beating the spacing tactic on CNN-BiGRU Ensemble models trained with WE-Random and Word2Vec (dynamic) embeddings. A bird's-eye view of the adversarial evaluation is illustrated in Appendix \ref{App:bird}.

To summarise, the spacing tactic proved to be the most devastating attack for both shallow and deep ML models, whereas EDA remained the least effective attack since all of the shallow ML and DL models show high robustness to this attack. The experiment shows the comparatively high robustness of word embedding to different adversarial attacks, especially paraphrasing and EDA, in making a better classification. Moreover, in comparison to shallow ML, the deep ML models demonstrated fair robustness to several adversarial tactics. In general, the results support the premise of the usefulness of contextual embedding and deep ML to resist different types of attacks. Moreover, it can be concluded that context plays a vital role in the classification and outcome of Spam detection as well as the robustness of the models. The classification accuracy of the models during adversarial evaluation with contextual embedding and deep ML remains comparatively high in comparison to the models trained with syntactic feature models and shallow ML. Furthermore, compared to deep ML, the classification performance of the ML models experienced a significant decrease in accuracy across most adversarial attacks. This indicates that ML models are more sensitive to adversarial attacks. Conversely, the classification performance of the deep ML models, especially the transformer-based architecture demonstrates good resistance to different sorts of adversarial attacks while classifying Spam messages.
\begin{table*}
\begin{center}
\caption{Robustness/ Adversarial Evaluation of Shallow ML Classifiers on The Holdout Split
}
\label{tab:robust-ml}
\newcolumntype{a}{>{\columncolor{yellow!50}}c}
\newcolumntype{b}{>{\columncolor{green!50}}c}
\newcolumntype{s}{>{\columncolor{blue!25}}c}
\tabcolsep=0.22cm
\renewcommand{\arraystretch}{1}
\scalebox{0.9}
{\begin{tabular}{ r | c | s | s | a | a | b | b | a | a | b | b | a | a | b | b}
 \hline
 \textbf{Feature}&  &
  \multicolumn{2}{c|}{\bf Actual} & 
  \multicolumn{2}{c|}{\bf Paraphrasing} &
  \multicolumn{2}{c|}{\bf EDA} &
  \multicolumn{2}{c|}{\bf Homograph} &
  \multicolumn{2}{c|}{\bf Spacing} &
  \multicolumn{2}{c|}{\bf Charswap} &
  \multicolumn{2}{c}{\bf Hybrid} \\
 \cline{3-16}
 \textbf{Model} & \textbf{Classifier} & \textbf{ACC} & \textbf{F1} & \textbf{ACC} & \textbf{F1} & \textbf{ACC} & \textbf{F1} & \textbf{ACC} & \textbf{F1} & \textbf{ACC} & \textbf{F1} & \textbf{ACC} & \textbf{F1} & \textbf{ACC} & \textbf{F1}\\ [0.25ex]
 \hline\hline
  \multirow{ 3}{*}{\makecell{Bow \\ (TF-IDF)}} & TCSVM & 80.9\% & 94\% & 51.1\% & 68\% & 79.1\% & 88\% & - & - &\cellcolor{red!75}\textbf{28.4\%} & \cellcolor{red!75}44\% & 35.6\% & 52\% & 29.8\% & 46\%\\
  & OCSVM & 65.8\% & 79\% & 22.6\% & 37\% & 62.2\% & 77\% & 63.6\% & 78\% & \cellcolor{red!75}\textbf{11.1\%} & \cellcolor{red!75}20\% & 19\% & 32\% & \cellcolor{red!75}\textbf{11.1\%} & \cellcolor{red!75}20\%\\
  & PU & 83.6\% & 91\% & 62.2\% & 77\% & 81.8\% & 90\% & 82.2\% & 90\% & \cellcolor{red!75}\textbf{32\%} & \cellcolor{red!75}48\% & 42.7\% & 60\% & 32.9\% & 49\%\\
  \hline
  \multirow{ 3}{*}{\makecell{Bigram \\ (TF-IDF)}} & TCSVM & 48\% & 65\% & 10.7\% & 19\% & 44.4\% & 62\% & - & - & 10.2\% & 19\% & \cellcolor{red!75}\textbf{5.8\%} & \cellcolor{red!75}11\% & 10.2\% & 19\%\\
  & OCSVM & 40.8\% & 58\% & \cellcolor{red!75}\textbf{26.7\%} & \cellcolor{red!75}42\% & 40\% & 57\% & 35.1\% & 52\% & 35.1\% & 52\% & 65.8\% & 79\% & 35.6\% & 52\%\\
  & PU & 60.4\% & 75\% & 20.9\% & 35\% & 52.9\% & 69\% & 56.4\% & 72\% & 13.3\% & 24\% & \cellcolor{red!75}\textbf{6.7\%} & \cellcolor{red!75}13\% & 12.9\% & 23\%\\
  \hline
  \multirow{ 3}{*}{\makecell{Trigram \\ (TF-IDF)}} & TCSVM & 28.9\% & 45\% & 1.3\% & 3\% & 20\% & 33\% & 25.3\% & 40\% & 1.8\% & 4\% & \cellcolor{red!75}\textbf{0\%} & \cellcolor{red!75}\textbf{0\%}& 1.8\% & 4\% \\
  & OCSVM & 66.2\% & 80\% & 80.4\% & 89\% & \cellcolor{red!75}\textbf{62.7\%} & \cellcolor{red!75}\textbf{77\%} & 64\% & \cellcolor{red!75}\textbf{78\%} & 76.4\% & 87\% & 96.4\% & 98\% & 76.9\% & 87\%\\
  & PU & 33.8\% & 50\% & 3.6\% & 7\% & 23.6\% & 38\% & 29.8\% & 46\% & 2.2\% & 4\% & \cellcolor{red!75}\textbf{0\%} & \cellcolor{red!75}\textbf{0\%}& 2.2\% & 4\%\\
  \hline
  \multirow{ 3}{*}{Word2Vec} & TCSVM & 83.1\% & 91\% & 76\% & 86\% & 83.1\% & 91\% & 33.3\% & 50\% & \cellcolor{red!75}\textbf{11.6\%} & \cellcolor{red!75}\textbf{21\%}& 39.6\% & 57\% & 17.3\% & 30\%\\
  & OCSVM & 87.5\% & 93\% & 88.9\% & 94\% & 89.8\% & 95\% & 74.4\% & 85\% & 85.3\% & 92\% & \cellcolor{red!75}\textbf{66.1\%} & \cellcolor{red!75}\textbf{80\%}& 90.7\% & 95\%\\
  & PU & 90.2\% & 95\% & 89.8\% & 95\% & 87.6\% & 93\% & 59.2\% & 74\% & 28\% & 44\% & \cellcolor{red!75}\textbf{0} & \cellcolor{red!75}0\% & 37.3\% & 54\%\\
  \hline
  \multirow{ 3}{*}{GloVe} & TCSVM & 79.6\% & 89\% & 69.8\% & 82\% & 78.2\% & 88\% & 72.0\% & 84\% & \cellcolor{red!75}\textbf{2.2\%} & \cellcolor{red!75}\textbf{4\%} & 40\% & 57\% & 7.1\% & 13\%\\
  & OCSVM & 91.6\% & 96\% & 94.2\% & 97\% & 92.8\% & 96\% & - & - & \cellcolor{red!75}\textbf{34.7\%} & \cellcolor{red!75}\textbf{51\%} & 67.1\% & 80\% & 54.7\% & 71\%\\
  & PU & 85.8\% & 92\% & 88\% & 94\% & 84.4\% & 92\% & 54.7\% & 71\% & \cellcolor{red!75}\textbf{8\%} & \cellcolor{red!75}\textbf{15\%} & 52.9\% & 69\% & 13.3\% & 24\%\\
  \hline 
{\makecell{fastText}} & fastText & 87.1\% & 93\% & 58.2\% & 74\% & 84.9\% & 92\% & 61.3\% & 76\% & \cellcolor{red!75}\textbf{37.8\%} & \cellcolor{red!75}55\% &  54.2\% & 70\% & 56.9\% & 73\%\\
  \hline
  \hline
\end{tabular}}
\end{center}
\end{table*}

\begin{table*}
\begin{center}
\caption{Robustness/ Adversarial Evaluation of Deep ML Classifiers on The Holdout Split.
}
\label{tab:robust-dl}
\newcolumntype{a}{>{\columncolor{yellow!50}}c}
\newcolumntype{b}{>{\columncolor{green!50}}c}
\newcolumntype{s}{>{\columncolor{blue!25}}c}
\tabcolsep=0.15cm
\renewcommand{\arraystretch}{1.1}
\scalebox{0.9}{\begin{tabular}{ r | c | s | s | a | a | b | b | a | a | b | b | a | a | b | b}
 \hline
 &  &
  \multicolumn{2}{c|}{\bf Actual} & 
  \multicolumn{2}{c|}{\bf Paraphrasing} &
  \multicolumn{2}{c|}{\bf EDA} &
  \multicolumn{2}{c|}{\bf Homograph} &
  \multicolumn{2}{c|}{\bf Spacing} &
  \multicolumn{2}{c|}{\bf Charswap} &
  \multicolumn{2}{c}{\bf Hybrid} \\
 \cline{3-16}
 \textbf{Classifier} & \textbf{Embedding} & \emph{ACC} &\emph{F1} &\emph{ACC} &\emph{F1} & \emph{ACC} &\emph{F1} &\emph{ACC} &\emph{F1} &\emph{ACC} &\emph{F1} &\emph{ACC} &\emph{F1} &\emph{ACC} &\emph{F1}\\ [0.25ex]
 \hline\hline
 {BERT} & bert-base-uncased & 89.8\% & 95\% & 77.8\% & 88\% & 88.4\% & 94\% & 68.4\% & 81\% & \cellcolor{red!75}\textbf{24.9\%} & \cellcolor{red!75}\textbf{40\%} & 69.8\% & 82\% & 40.0\% & 57\% \\
  \hline
   {ELMo} & ELMo & 89.3\% & 94\% & 77.7\% & 88\% & 87.9\% & 94\% & 83.9\% & 91\% & \cellcolor{red!75}\textbf{44.6\%} & \cellcolor{red!75}\textbf{62\%} & 82.1\% & 90\% & 55.4\% & 71\% \\
  \hline
   {RoBERTa} & roberta-base & 92.9\% & 96\% & 83.6\% & 91\% & 92.4\% & 96\% & - & - & \cellcolor{red!75}\textbf{51.6}\% & \cellcolor{red!75}\textbf{68\%} & 89.3\% & 94\% & 81.3\% & 90\% \\
  \hline
   {XLM-RoBERTa} & xlm-roberta-base & 90.2\% & 95\% & 78.2\% & 88\% & 89.8\% & 95\% & 91.1\% & 95\% & \cellcolor{red!75}\textbf{34.2\%} & \cellcolor{red!75}\textbf{51\%} & 83.1\% & 91\% & 54.2\% & 70\% \\
  \hline
   {DistilBERT} & distilbert-base-uncased & 95.1\% & 97\% & 83.6\% & 91\% & 92.9\% & 96\% & 78.7\% & 88\% & \cellcolor{red!75}\textbf{43.1\%} & \cellcolor{red!75}\textbf{60\%} & 76.0\% & 86\% & 63.1\% & 77\% \\
  \hline
   {LSTM} & WE-Random & 75.1\% & 86\% & 53.3\% & 70\% & 72.4\% & 84\% & 42.2\% & 59\% & \cellcolor{red!75}\textbf{9.8\%} & \cellcolor{red!75}\textbf{18\%} & 56.0\% & 72\% & 14.7\% & 26\% \\
  \hline
   {BiLSTM} & WE-Random & 70.2\% & 83\% & 48.4\% & 65\% & 68.0\% & 81\% & 28.7\% & 45\% & \cellcolor{red!75}\textbf{8.4\%} & \cellcolor{red!75}\textbf{16\%} & 44.9\% & 62\% & 11.1\% & 20\% \\
  \hline
  \multirow{ 2}{*}{CNN} & WE-Random & 74.2\% & 85\% & 50.7\% & 67\% & 70.2\% & 83\% & 34.1\% & 51\% & \cellcolor{red!75}\textbf{7.6}\% & \cellcolor{red!75}\textbf{14}\% & 50.7\% & 67\% & 12.4\% & 22\% \\
  & GloVe (static) & 78.7\% & 88\% & 54.7\% & 71\% & 77.3\% & 87\% & 42.2\% & 59\% & \cellcolor{red!75}\textbf{7.1\%} & \cellcolor{red!75}\textbf{13\%} & 56.9\% & 73\% & 11.1\% & 20\% \\
  \hline
  \multirow{ 3}{*}{TCN} & TCN & 80.9\% & 89\% & 68.4\% & 81\% & 80.4\% & 89\% & 60.9\% & 76\% & \cellcolor{red!75}\textbf{24.0\%} & \cellcolor{red!75}\textbf{39\%} & 55.1\% & 71\% & 17.8\% & 30\% \\
  & Word2Vec (static) & 83.1\% & 91\% & 73.8\% & 85\% & 82.2\% & 90\% & 72.0\% & 84\% & \cellcolor{red!75}\textbf{15.1\%} & \cellcolor{red!75}\textbf{26\%} & 76.9\% & 87\% & 12.4\% & 22\% \\
  & Word2Vec (dynamic) & 91.1\% & 95\% & 80.4\% & 89\% & 90.7\% & 95\% & 87.6\% & 93\% & \cellcolor{red!75}\textbf{42.2}\% & \cellcolor{red!75}\textbf{59\%} & 88.9\% & 94\% & 43.6\% & 61\% \\
  \hline
  \multirow{ 3}{*}{\makecell{Ensemble \\ (CNN-BiGRU)}} & WE-Random & 79.1\% & 88\% & 68.0\% & 81\% & 79.1\% & 88\% & 58.6\% & 74\% & 12.4\% & 22\% & 64.9\% & 79\% & \cellcolor{red!75}\textbf{9.3\%} & \cellcolor{red!75}\textbf{17\%} \\
  & Word2Vec (static) & 87.6\% & 93\% & 85.8\% & 92\% & 85.8\% & 92\% & 84.0\% & 91\% & \cellcolor{red!75}\textbf{12.4}\% & \cellcolor{red!75}\textbf{22\%} & 87.6\% & 93\% & 16.0\% & 28\%\\
  & Word2Vec (dynamic) & 84.9\% & 92\% & 79.1\% & 88\% & 85.8\% & 92\% & 65.3\% & 79\% & 27.6\% & 43\% & 69.8\% & 82\% & \cellcolor{red!75}\textbf{24.4\%} & \cellcolor{red!75}\textbf{39\%}\\
  \hline 
  \hline
\end{tabular}}
\end{center}
\end{table*}
\section{Discussion \& Future Research Direction}
In this section, we summarise the key observations that we found while comprehensively evaluating the ML models concerning performance and robustness.

\textbf{Spam detection.} In the first set of experiments, we trained thirty-one ML models to evaluate and identify suitable models for SMS Spam detection. While the majority of shallow ML models provide excellent accuracy, just one model achieved a decent f1-score, indicating that the shallow ML models have a poor Spam detection rate. In contrast, a few deep ML models achieve acceptable f1-scores when compared to shallow ML models. This demonstrates the superiority of deep ML over shallow ML in SMS Spam classification.

\textbf{Spam detection with few positive samples (small training set problem).} 
The PU learning method learns from the positive cases in the data and applies what it has learned to filter Spam messages. The PU technique outperforms OCSVM in all feature models and gets a high f1-score and accuracy in both performance and adversarial assessment, which is similar to TCSVM and fastText. This makes it a good alternative when only positive samples and small datasets are available.

\textbf{Tackling the imbalance dataset.} Currently, all of the public Spam SMS datasets, including the super dataset, are highly imbalanced, therefore it is not suitable to apply traditional ML techniques. As such, it is surprising that imbalanced classification in Spam SMS detection does not get more attention. The OCSVM (one-class learning) in this research cannot perform well, however, it is strongly recommended to explore other approaches like one-class neural networks, etc.

\textbf{Lack of robustness.} One of the most important aspects of this research was to explore how adversarial attacks can be used to target shallow ML and deep ML-based SMS Spam models by generating adversarial samples. Therefore, a second set of experiments was designed for this purpose, and the previously trained models were assessed for robustness using different adversarial tactics. Overall, both the shallow ML and deep ML models had a significant decrease in classification performance when adversarial samples were present. The adversarial evaluation proved that all ML models are currently vulnerable to a wide range of adversarial attacks. Therefore, a robust model for SMS Spam detection with high-performance accuracy is required.

\textbf{Lack of cross evaluation.} Common techniques for training and testing the data involve k-fold cross-validation or splitting the original dataset into training (usually 80\%-70\%) and testing (usually 20\%-30\%) data. But in such cases, the source of the data remains the same and the test set may have attributes that are common to the training set—thus ensuring better performance of the classifier. This, however, may not be the case in a real-world scenario. Therefore, the proposed methodology should be cross-evaluated (trained on one dataset and tested on a different dataset). The results received in Table \ref{tab:robust-ml} and Table \ref{tab:robust-dl} from unaltered instances of holdout split on the same models are quite low in comparison to the results in Table \ref{tab:perform-ml} and Table \ref{tab:perform-dl}, which clearly prove this point.

\textbf{Monitoring the threat landscape.} For effective Spam filtering, it is mandatory to stay on top of adversaries by proactively looking for new threats that may arise due to new technologies like the internet of things or current trends and events (like COVID-19). It is recommended to search and crawl the latest threats from social media (Twitter etc) and scam observatories.

\textbf{Lack of generalization abilities and adversarial training of models.} The newly developed Punycode attack for SMS Spam proved to be effective in evading different models by encapsulating the Spam keywords. However, one of the most important use cases of Punycode is to make phishing URLs look like legitimate website URLs, which needs to be evaluated in detail for SMiShing. Moreover, research work needs to be conducted to tackle the fact that SMS Spam filtering is a concept drift problem. Researchers have been working towards this goal of designing a general method. However, none of the existing works meet this need. Therefore, future work should focus more on designing a general defense for a single NLP task and then extending it to other NLP tasks. Adversarial training of ML models is one important aspect to consider for this purpose. Recent adversarial examples can be leveraged to improve the performance and robustness of ML models. Also, we believe that a blend of encoding models would be a plausible solution to this problem.

\textbf{Performance, robustness, cost, and time trade-off.} DL models are powerful and have the capacity to solve sophisticated problems by exploiting the computational power of modern GPUs. The performance assessment of shallow and deep ML models in Table \ref{tab:perform-ml} and Table \ref{tab:perform-dl} respectively clearly proved the dominance of deep ML over shallow ML (high f1-score of several DL models). Moreover, the results received from the robustness assessment of the models indicate that transformer-based DL models can significantly improve model robustness. DL models are slow to train and require a lot of computational power. The shallow ML model, fastText, was much faster than DL for training and evaluation and scored well in Spam classification (f1-score of 82\%) and fairly in robustness assessment also in comparison to other ML models (performance a little lower than DL models). Therefore, fastText is a viable alternative to DL models in situations where low computation and training time are desired.

\textbf{Word embedding as default feature extractor.} The results received in both the performance evaluation and robustness assessment demonstrated the importance of context-based word embedding as a default feature extractor. This is one potential step to counter adversarial attacks along with other techniques such as adversarial training. Any model constructed on top of semantic-based word embedding performs well in terms of Spam classification and robustness.

\textbf{Appropriate Evaluation metrics.} From Table \ref{tab:perform-ml}\&\ref{tab:perform-dl}, we concluded that the accuracy metric does not capture the true performance of the classifier in the case of unbalanced data, and currently, all of the public Spam SMS datasets are highly imbalanced. A better way to evaluate would be to include the F1 score along with the accuracy metric.

\textbf{Requirement of Collaborative approach.} From the literature, we realised that very few studies were performed on the server side (network providers) due to the poor collaboration of researchers with the industry. Therefore, hybrid techniques that blend content-based and non-content-based aspects with a collaborative approach (solutions that reside both on the network provider’s side and the user's side) may also need to be investigated. The non-content-based techniques make use of particular message attributes like the message's size or temporal and network features. This might not only improve the performance but also overcome some of the challenges faced in Spam SMS detection, such as limited textual content, data sparsity, and unstandardized acronyms.

\textbf{Crowdsourcing Spam SMS collection.} We recommend a crowdsourcing approach to get the most Spam data, reduce the problem of data imbalance, increase data diversity, and cover the most recent Spam campaigns that are targeting mobile users. Note that contemporary NLP models have a huge number of parameters, so training them without a large dataset is not ideal.

\textbf{Lack of better unsupervised ML methods.} We observed an important gap in the detection techniques employed for Spam SMS. The majority of the implementations use supervised learning methods. The major necessity for training and evaluating the performance of such supervised systems is the considerable amount of labelled data, which brings us back to a major security challenge – data availability. Therefore, we implemented the OCSVM for Spam detection in this work, however, more efficient unsupervised approaches need to be evaluated for this problem.
\section{Conclusion}
In this study, we presented an accumulated view of the performance and robustness of a large number of AI models in detecting spam SMS. We evaluated and compared the performance of AI models on a large new SMS dataset to determine their efficacy in the evolving threat landscape. We found these models to be highly effective in identifying the Ham SMS. However, only a few deep learning models perform well in classifying Spam with a precision score above 90\% whereas others fail to achieve this landmark. Next, we designed several black-box attack strategies using different levels of adversarial perturbations, keeping the present threat landscape in mind, which significantly lowered the effectiveness of all trained models in our large-scale experiments. In addition, we highlighted several challenges and discussed future research prospects. We anticipate that this work will serve as a catalyst for qualitative research leading to the development of a robust Spam SMS detection model.

\bibliographystyle{IEEEtran}
\bibliography{main.bib}

\section*{Acknowledgement}
The work was done while holding a joint PhD scholarship sponsored by Macquaire University, Sydney and Higher Education Commission of Pakistan.

\appendix
\section{Appendix}
\subsection{\textbf{Spam Labelling Rules}}\label{App:rules}
We derived a set of 8 rules to govern our manual labelling of SMS as Spam or Ham.
The guidelines were developed after a thorough analysis of labelled SMS corpora, debate among the authors and review of different scam types on scam observatories (scam watch, Action Fraud). A single expert analysed the unlabeled SMSes in the consolidated dataset and labelled them using the defined labelling criteria. In the case of two or more reviewers, there may be differences in judging the intent of SMSes. The Spam labelling rules we utilised are summarised in Table \ref{tab:rules}.

\begin{table}
\begin{center}
\caption{Spam Labelling Rules}
\label{tab:rules}
\def\arraystretch{1}%
{\begin{tabular}{p{0.75cm}|p{7cm}} 
 \hline
  {\bf Rule} & {\bf Description} \\
 \hline\hline
  Rule1 & Promotional or unwanted messages (advertising, proselytizing etc)\\ 
\hline 
Rule2 & Messages with users to visit or tap on the links within message \\
\hline
Rule3 & Asking users to contact on email within text message \\
\hline
Rule4 & Asking users to contact back on the same number or another contact number within text message.\\
\hline
Rule5 & Asking users for personal or sensitive information \\
\hline
Rule6 & Asking or requesting users for payment \\
\hline
Rule7 & Asking users to forward or circulate the message \\
\hline
Rule8 & Asking users to download or install a file \\
 \hline
 \hline
\end{tabular}}
\end{center}
\end{table}

\subsection{Adversarial Evaluation of ML Models}\label{App:bird}
\begin{figure}
     \centering
     \begin{subfigure}[b]{0.45\textwidth}
         \centering
          \includegraphics[width=\textwidth]{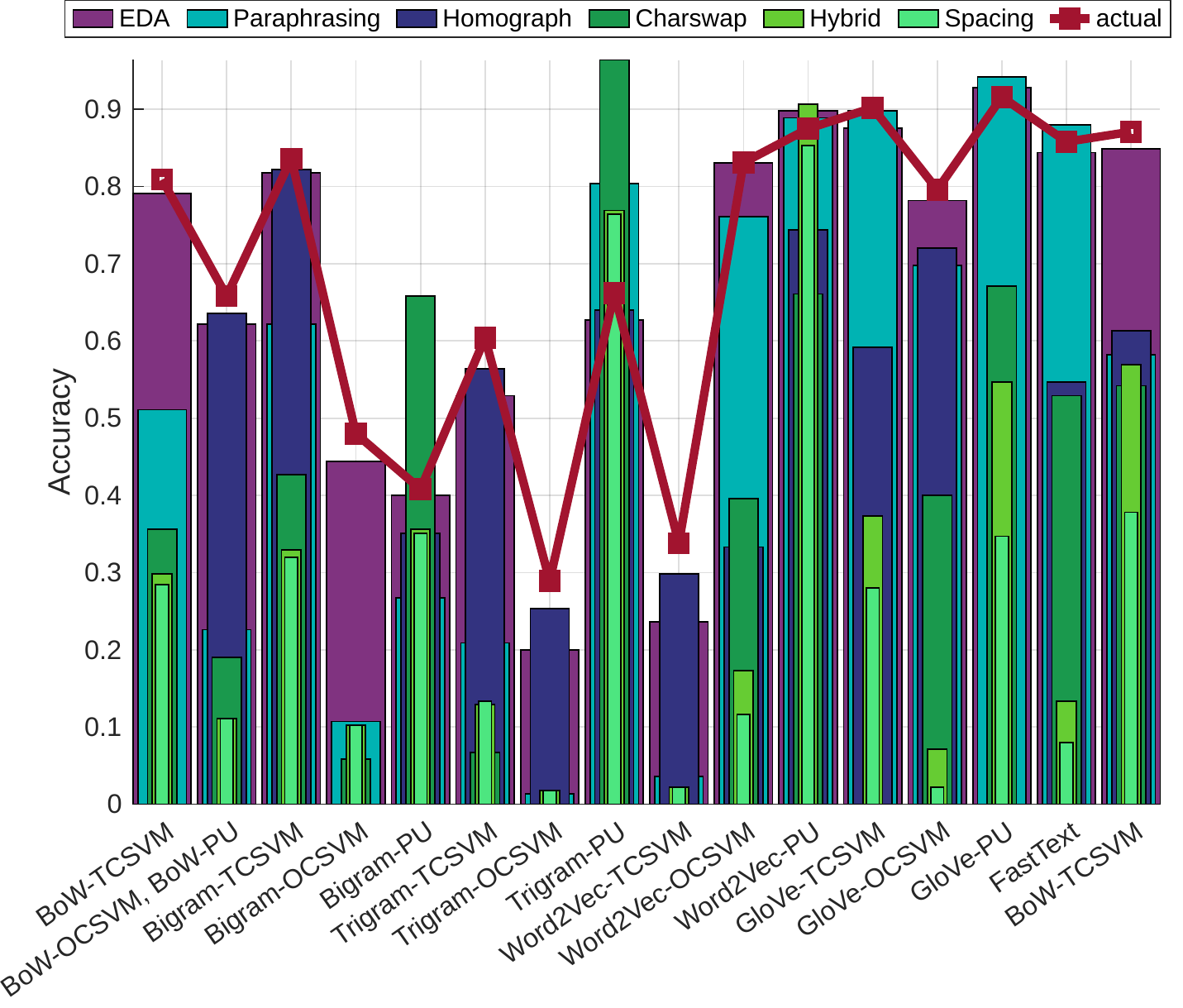}
         \caption{Shallow ML Models.}
     \end{subfigure}
     \begin{subfigure}[b]{0.45\textwidth}
         \centering
         \includegraphics[width=\textwidth]{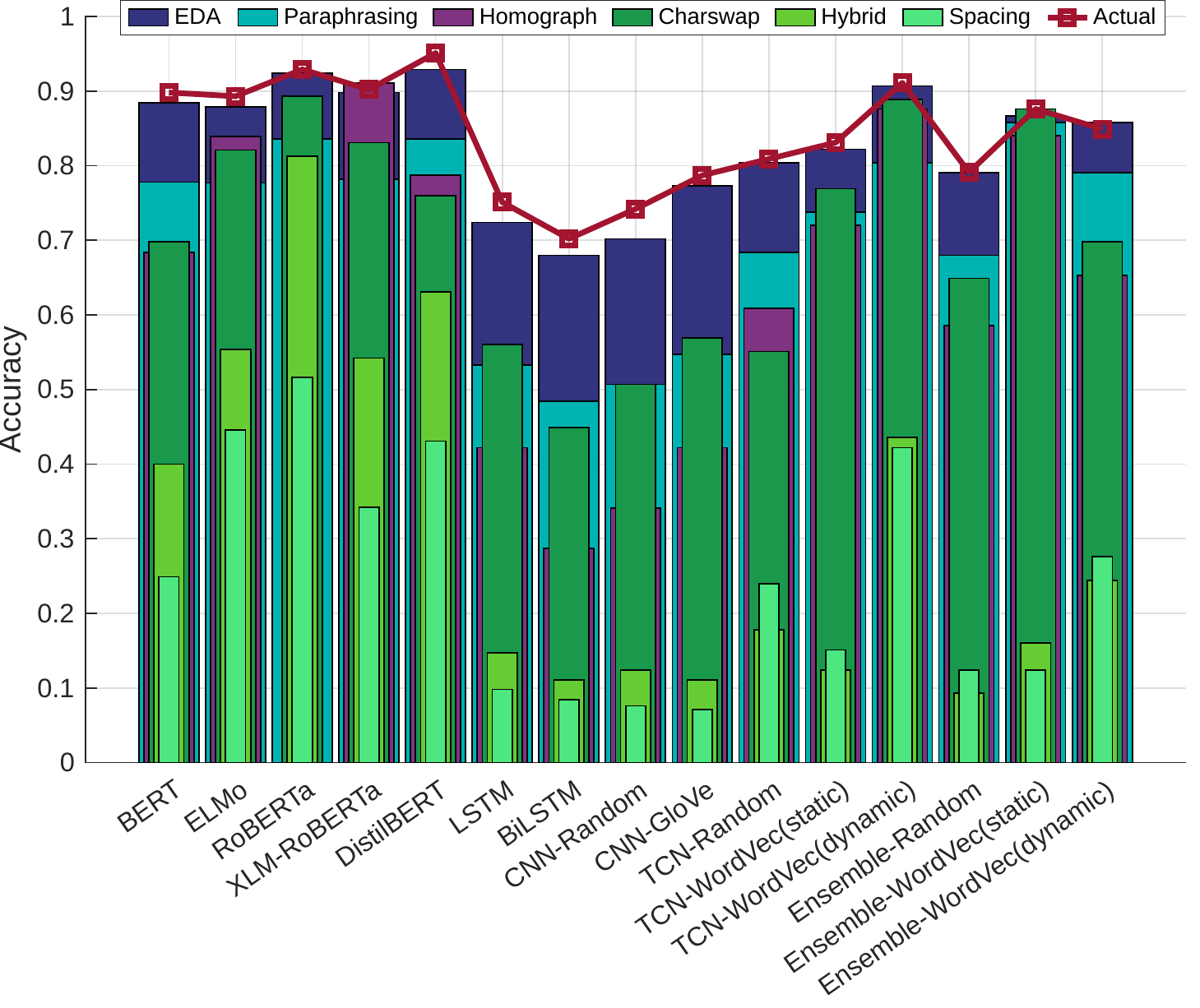}
         \caption{DL Models.}
     \end{subfigure}
     \hfill
        \caption{A bird's-eye view of the adversarial evaluation of ML models. The results of different attacks against each model are stacked together.}\label{fig:birdeye}
\end{figure}

As discussed in Section \ref{sec:aml}, Adversarial Evaluation of ML Models reveals that all of them are vulnerable to some type of attack. In most adversarial techniques, the classification accuracy of ML models was poor to that of DL models (see Figure \ref{fig:birdeye}). 
\subsection{Explanations}\label{app:explan}
\textbf{Word2Vec} uses a neural network to learn word associations from large corpora and generate a vector representation for words or tokens and captures semantic similarity between words. Word2Vec creates a vector space with each unique word in the corpus, often with several hundred dimensions, such that words with similar contexts in the corpus are close to one another in the space. Moreover, the vectors for unknown words are randomly initialized using a generic normal distribution.

\textbf{BERT} is a language model developed by Google. Generally, language models read the input sequence in one direction: either left to right or right to left. This kind of one-directional training works well when the aim is to predict or generate the next word. But in order to have a deeper sense of language context, BERT uses bidirectional training. BERT applies the bidirectional training of Transformer to language modeling and learns the text representations. BERT generates a language model depending on its context, capturing lexical, semantic and grammatical features. BERT is designed to pre-train deep bidirectional representations from unlabeled text. Consequently, the pre-trained BERT model is used for extracting word embedding/ representations. 
\textbf{LSTM} Due to the vanishing gradient issue, simple RNNs are incapable of representing longer contextual connections. They have mostly been supplanted by so-called long short-term neural networks (LSTMs), which are similar to RNNs but can capture the lengthier context included in texts~\cite{hochreiter1997long}.

\textbf{Bidirectional LSTM (BiLSTM)} LSTMs can only process unidirectional sequences. Consequently,  state-of-the-art techniques based on LSTMs evolved into so-called bidirectional LSTMs that can read the context left to right as well as right to left~\cite{graves2005framewise}.

\textbf{Bidirectional Gated Recurrent Unit (BiGRU)} Bidirectional GRU’s ~\cite{qing2019novel} are a type of bidirectional recurrent neural networks with only the input and forget gates. It enables the utilisation of knowledge from prior and subsequent time steps to create predictions about the present state.

\textbf{DistilBERT} The DistilBERT \cite{sanh2019distilbert} is a lighter, faster, and more affordable version of BERT. It learns a distilled (approximate) version of BERT that retains 95\% of the performance but uses half the parameters. Specifically, it lacks token-type embeddings, a pooler, and preserves just half of Google's BERT's layers. DistilBERT makes use of a method called distillation to resemble Google's BERT, i.e. replacing the huge neural network with a smaller one. Following training a big neural network, the network's whole output distributions can be approximated using a smaller network.

\textbf{Robustly optimized BERT approach (RoBERTa)} RoBERTa \cite{liu2019roberta} is a retraining of BERT that incorporates an enhanced training process, 1,000\% more data, and significantly more computational capacity. To improve the training approach, RoBERTa omits the Next Sentence Prediction (NSP) task from BERT's pre-training and introduces dynamic masking, in which the masked token fluctuates between training epochs. Additionally, it was demonstrated that bigger batch sizes were more effective during the training process.

\textbf{Cross-lingual Language Model (XLM)} XLM is an enhanced version of BERT suggested by Facebook AI \cite{conneau2019cross} to deliver state-of-the-art results in a variety of natural language processing tasks, most notably cross-lingual classification, supervised and unsupervised machine translation. XLM employs a dual-language training mechanism in conjunction with BERT to acquire knowledge about the relationships between words in multiple languages. When a pre-trained model is used to initialize the translation model, the model outperforms other models in a cross-lingual classification problem and considerably improves machine translation.

\end{document}